\documentclass[prl,twocolumn,amsmath,showpacs]{revtex4}

\usepackage{graphicx}
\usepackage{dcolumn}
\usepackage{bm}

\begin{document}

\title{Nonlinear envelope equation for broadband optical pulses in quadratic media}

\author{Matteo Conforti, Fabio Baronio and Costantino De Angelis}

\affiliation{ CNISM and Dipartimento di Ingegneria dell'Informazione, Universit\`a di Brescia, \\
Via Branze 38, 25123 Brescia, Italy
}%

\date{\today}

\begin{abstract}
\textbf{We derive a nonlinear envelope equation to describe the
propagation of broadband optical pulses in second order nonlinear
materials. The equation is first order in the propagation
coordinate and is valid for arbitrarily wide pulse bandwidth. Our
approach goes beyond the usual coupled wave description of
$\chi^{(2)}$ phenomena and provides an accurate modelling of the
evolution of ultra-broadband pulses also when the separation into different coupled frequency components is not possible or not profitable. }

\end{abstract}

\maketitle

The analysis of optical pulse propagation typically involves the
definition of a complex envelope whose variation is supposed to be
``slow'' with respect to the oscillation of a carrier frequency
(``slowly varying envelope approximation'', SVEA \cite{Boyd}). In
the frequency domain this assumption is equivalent to require that
the bandwidth of the envelope is narrow with respect to the
carrier frequency. Different works showed that it is possible to
extend the validity of a proper generalization of the envelope
equation (for example the ``Nonlinear Envelope Equatio'' (NEE) of
Brabec and Krausz) to pulse duration down to the single optical
oscillation cycle
\cite{Brabec97,Geissler99,Brabec00,Housakou01,Kolesik02} and to
the generation of high order harmonics \cite{Genty07}. When second
order nonlinearities are considered, the usual approach is to
write coupled equations for the separated frequency bands relevant
for the process \cite{Kinsler03,Moses06}. However when
ultra-broadband $\chi^{(2)}$ phenomena take place, the different
frequency bands might merge, generating a single broad spectrum,
as observed in recent experiments \cite{Langrock07}. Obviously in
these cases the coupled NEE description of the propagation fails
due to the overlapping between different frequency bands.

The scope of this Letter is to  provide a single wave envelope
equation to describe ultra-broadband $\chi^{(2)}$ interactions. To
date, such a model is not available and the only way to
numerically describe phenomena as those reported in Ref.
\cite{Langrock07} is to solve directly Maxwell equations in time
domain, with an immense computational burden. Our equation,
besides providing a powerful tool for analytical treatment due to
its simplicity, can be easily solved with a modest computational
effort and can be easily generalized to include other kind of
nonlinearities such as Kerr or Raman.

As far the linear dispersive terms are concerned, our derivation of the envelope equation builds upon the work of Brabec and Krausz
\cite{Brabec97}, that carry to a simple model that was shown
(theoretically and experimentally) to be accurate in most
situations. Starting from Maxwell equations (written in MKS
units), neglecting transverse dimensions (i.e considering the
propagation of plane waves), we can obtain the 1+1D wave equation
for the electric field $E(z,t)$:
\begin{eqnarray}
\nonumber \frac{\partial^2 E(z,t)}{\partial z^2
}-\frac{1}{c^2}\frac{\partial^2}{\partial
t^2}\int_{-\infty}^{+\infty}E(z,t')\varepsilon(t-t')dt'\\=\frac{1}{\varepsilon_0
c^2}\frac{\partial^2}{\partial t^2} P_{NL}(z,t),
\end{eqnarray}
that can be written in frequency domain, by defining the Fourier
transform $\mathcal{F}[E](\omega)=\hat
E(\omega)=\int_{-\infty}^{+\infty} E(t)e^{-i\omega t}dt$:
\begin{equation}\label{onde_freq}
\frac{\partial^2 \hat E(z,\omega)}{\partial z^2} +
\frac{\omega^2}{c^2}\hat \varepsilon(\omega)\hat
E(z,\omega)=-\frac{\omega^2}{\varepsilon_0c^2}\hat
P_{NL}(z,\omega),
\end{equation}
where $c$ is the vacuum velocity of light, $\varepsilon_0$ is the
vacuum dielectric permittivity,
$\hat\varepsilon(\omega)=1+\hat\chi(\omega)$ and
$\hat\chi(\omega)$ is the linear electric susceptibility.

We consider now the electric field $E$ and the nonlinear
polarization $P_{NL}$ as the product of a complex envelope  and a
carrier wave: $E(z,t)=A(z,t)/2e^{i\omega_0t-i\beta_0z} + c.c.$,
$P_{NL}(z,t)=A_p(z,t)/2e^{i\omega_0t-i\beta_0z} + c.c$ [in
frequency domain reads: $\hat E(z,\omega)=\hat
A(z,\omega-\omega_0)/2e^{-i\beta_0z} + \hat
A^*(z,-\omega-\omega_0)/2e^{i\beta_0z}$], where $\omega_0$ is a
reference frequency, $\beta_0=Re[k(\omega_0)]$ and
$k(\omega)=(\omega/c)\sqrt{\hat\varepsilon(\omega)}$ is the
 propagation constant. \\
Particular care must be devoted to the definition of the complex
envelope, since we do not want to put any limitation to the
frequency extent of the signals. This aspect is commonly
overlooked in literature, and it is taken for granted that the
band of the envelope is ``narrow'' in some sense. We shall see
later that for quadratically nonlinear media, a proper definition
of the envelope is crucial. As usual in the theory of modulation
\cite{Haykin}, we define the analytic representation of the
electric field:
\begin{equation}\label{analytic}
\tilde E(z,t)= E(z,t)+ i \mathcal{H}[E](z,t),
\end{equation}
where
\begin{equation}\mathcal{H}[E](z,t)=\frac{1}{\pi}p.v.\int_{-\infty}^{+\infty}\frac{E(z,t')}{t-t'}dt'\end{equation}
is the Hilbert transform of the electric field ($p.v.$ indicates
the Cauchy principal value of the integral). The Fourier transform
of the analytic signal reads:
\begin{equation}
\hat{\tilde E}(z,\omega)=\left\{ \begin{array}{lll}
2\hat E(z,\omega) & \textrm{if $\omega>0$}\\
\hat E(z,0) & \textrm{if $\omega=0$} \\
0 & \textrm{if $\omega<0$}
\end{array} \right. ,
\end{equation}
that is a signal that contains only the positive frequency content
of the electric field. Due to reality of $E(z,t)$, its Fourier
transform has Hermitian symmetry, so that only the positive (or the negative)
frequencies carry information, and we can write:
\begin{equation}
\hat E(z,\omega)=\frac{1}{2}\hat{\tilde E}(z,\omega) +
\frac{1}{2}\hat{\tilde E}^*(z,-\omega),
\end{equation}
and eventually we can define the complex electric field envelope
as:
\begin{equation}\label{envelope}
A(z,t)=\tilde E(z,t)e^{-i\omega_0t+i\beta_0z},
\end{equation}
i.e. the inverse Fourier transform of the positive frequency
content of $E$ shifted towards the low frequency part of the
spectrum by an amount $\omega_0$.\emph{ It is worth noting that no
approximations on the frequency extent of the envelope has been
done}, and so $\textrm{ supp} \{\hat A(z,\omega)
\}=(-\omega_0,+\infty)$.

The substitution of expressions of $\hat E(z,\omega)$ and $\hat
P_{NL}(z,\omega)$ in Eq. (\ref{onde_freq}), Taylor-expansion of
$k(\omega)$ about $\omega_0$, application the \emph{slowly
evolving wave approximation} (SEWA, that is the neglect of second
space derivative in the coordinate system moving at the group
velocity at the reference frequency), followed by an inverse
Fourier transform yields \cite{Boyd,Brabec97,Brabec00}:
\begin{equation}
\frac{\partial A(z',\tau)}{\partial z'} + i D
A(z',\tau)=-i\frac{\omega_0^2}{2\beta_0c^2\varepsilon_0}\bigg(1-\frac{i}{\omega_0}\frac{\partial}{\partial
\tau }\bigg)A_p(z',\tau),
\label{brabec}
\end{equation}
where
$D=\sum_{m=2}^{\infty}\frac{1}{m!}k_m(-i\frac{\partial}{\partial
t})^m$, $k_m=\frac{\partial k}{\partial\omega}(\omega_0)$, $z'=z$
and $\tau=t-k_1z$ is the coordinate system  moving at the
reference group velocity. \\
It is worth noting that when the requirement
$|\frac{\beta_0-\omega_0k_1}{\beta_0}|<<1$ is accomplished, SEWA
does not explicitly impose a limitation on pulse duration and
bandwidth. Far from resonances, this requirement is fulfilled in
the majority of parametric processes in which all waves propagate
in the same direction.

We now consider an instantaneous second order $\chi^{(2)}$
nonlinearity, giving rise to the following nonlinear
polarization:
\begin{eqnarray}\label{Pnl}
& & P_{NL}(z,t)=\varepsilon_0\chi^{(2)}E(z,t)^2\\
\nonumber& &=\varepsilon_0\chi^{(2)}Re[A(z,t)e^{i\omega_0t-i\beta_0z}]^2\\
\nonumber& &=
\frac{\varepsilon_0\chi^{(2)}}{4}\bigg[A^2e^{2i\omega_0t-2i\beta_0z}+A^{*2}e^{-2i\omega_0t+2i\beta_0z}+2|A|^2\bigg].
\end{eqnarray}
It is worth noting that, due to the definition of $A$, the first
(second) term in the square brackets contains only positive
(negative) frequencies, whereas the third has both. It is now
apparent that it is impossible to separate the nonlinear
polarization in two distinct and ``narrow'' bands for the positive
and negative frequencies, as common in cubic media. Moreover the
neglect of the third term leads to totally wrong results
 (this term is responsible for difference frequency generation). \\
By going through the steps (\ref{analytic})-(\ref{envelope}) we
can instead correctly define the nonlinear polarization envelope:
\begin{eqnarray} \label{Ap_ex}
& &A_p(z,t)=\tilde
P_{NL}(z,t)e^{-i\omega_0t+i\beta_0z}\\
 & &=
\frac{\varepsilon_0\chi^{(2)}}{2}\bigg[A^2e^{i\omega_0t-i\beta_0z}+\bigg(|A|^2+i\mathcal{H}[|A^2|]\bigg)e^{-i\omega_0t+i\beta_0z}\bigg] \nonumber
\end{eqnarray}
Before inserting Eq. (\ref{Ap_ex}) into Eq. (\ref{brabec}), the
term $|A|^2$ in Eqs. (\ref{Pnl}) and (\ref{Ap_ex}) deserves
further comments, since it is centered around zero in frequency
domain. In particular to obtain the nonlinear polarization
envelope in Eq. (\ref{Ap_ex}) we had to filter out the negative
frequency components of $\hat P_{NL}(\omega)$, as done for $\hat
E(\omega)$. We not however that (i) $\hat A(z,\omega-\omega_0)$
does not contain negative frequency by definition, (ii) $P_{NL}$
is a small perturbation to linear polarization and (iii) negative
frequencies cannot be phase-matched. It follows that the task of
filtering the negative frequency components of $|A|^2$ can be left
to the propagation equation instead of having it explicitly in the
definition of $A_p(z,t)$. In other words, when inserting Eq.
(\ref{Ap_ex}) into Eq. (\ref{brabec}), we can write:
$|A|^2+i\mathcal{H}[|A^2|]\approx2|A|^2$.  We have checked
numerically the good accuracy of this approximation. Even if this
approximation in not necessary in the numerical solution (it is
straightforward to calculate the exact nonlinear polarization
envelope in frequency domain), it is suitable to obtain a simple
and manageable model for further analytical investigations.

The NEE for $A=A(z',\tau)$ becomes
\begin{eqnarray}\label{NEE2}
\nonumber \frac{\partial A}{\partial z'} + i D
A=-i\frac{\chi^{(2)}\omega_0^2}{4\beta_0c^2}\bigg(1-\frac{i}{\omega_0}\frac{\partial}{\partial
\tau
}\bigg)\bigg[A^2e^{i\omega_0t-i(\beta_0-k_1\omega_0)z}\\
+2|A|^2e^{-i\omega_0t+i(\beta_0-k_1\omega_0)z}\bigg],
\end{eqnarray}
or, performing derivatives,
\begin{eqnarray}\label{NEE2_der}
\nonumber & &\frac{\partial A}{\partial z'} + i D
A=\\
\nonumber & &
-i\frac{\chi^{(2)}\omega_0^2}{4\beta_0c^2}\bigg[\bigg(2A^2-\frac{2i}{\omega_0}A\frac{\partial
A}{\partial\tau}\bigg)e^{i\omega_0t-i(\beta_0-k_1\omega_0)z}\\
& & -\frac{4i}{\omega_0}Re\bigg[ A^*\frac{\partial A}{\partial
\tau} \bigg]e^{-i\omega_0t+i(\beta_0-k_1\omega_0)z}\bigg].
\end{eqnarray}
Equation (\ref{NEE2}) or (\ref{NEE2_der}) constitutes the main
result of this Letter. This nonlinear envelope equation first
order in propagation coordinate provides a powerful means of
describing light pulse propagation in dispersive quadratically
nonlinear media.

Starting from Eq. (\ref{NEE2}) it is straightforward to show that
our equation conserves the total energy of the field, i.e.
$\frac{d}{dz'}\int_{-\infty}^{+\infty}|A(z',\tau)|^2d\tau=0$. It
can also be shown that the total energy is conserved even if the
non approximated nonlinear polarization envelope $A_p$
[Eq.(\ref{Ap_ex})] is used.

We solved Eq. (\ref{NEE2}) by split-step Fourier method exploiting fourth
order Runge-Kutta scheme for the nonlinear step.
\begin{figure}[h]
  \centering
      \includegraphics[width=0.35\textwidth]{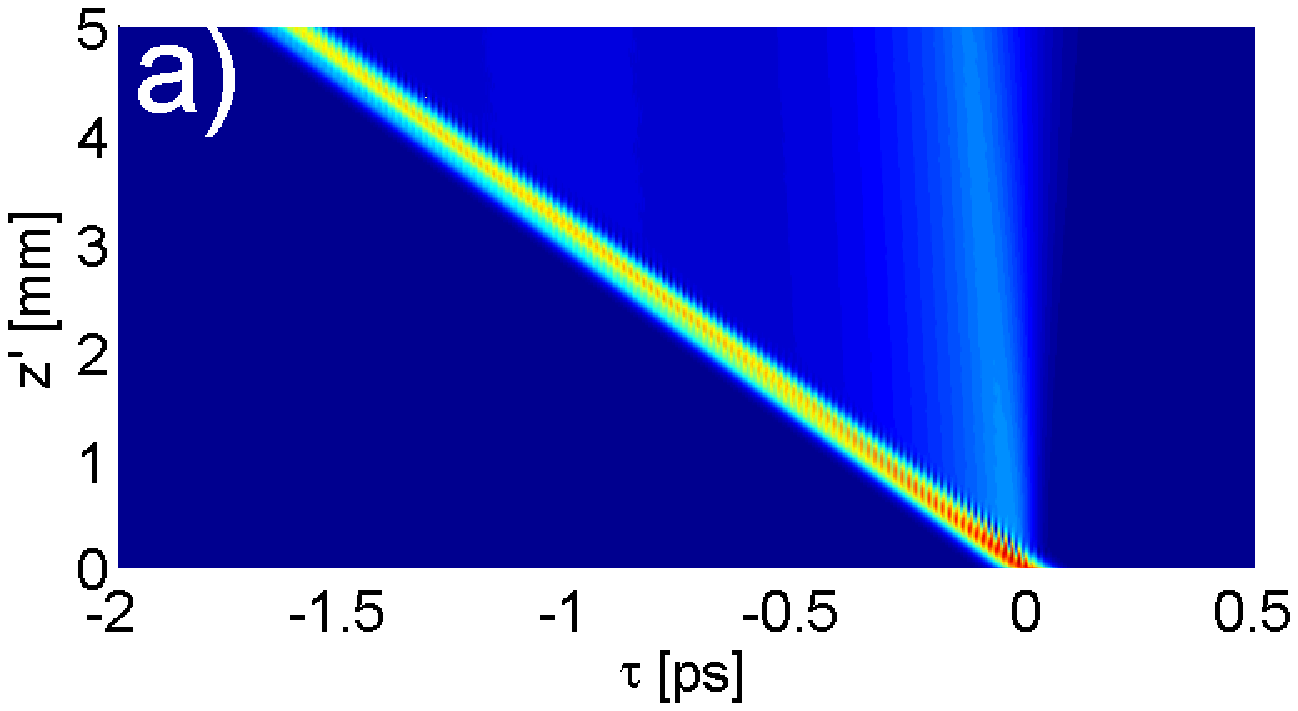}
      \includegraphics[width=0.35\textwidth]{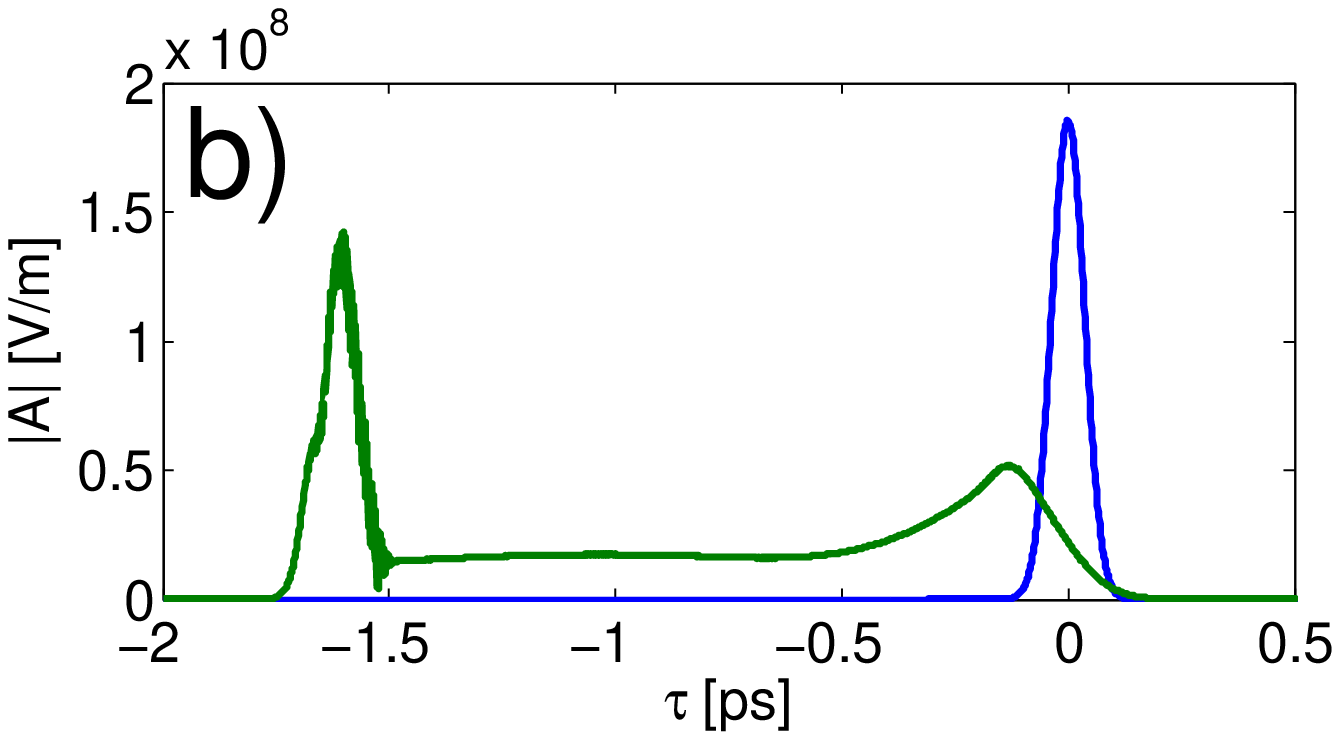}
\includegraphics[width=0.35\textwidth]{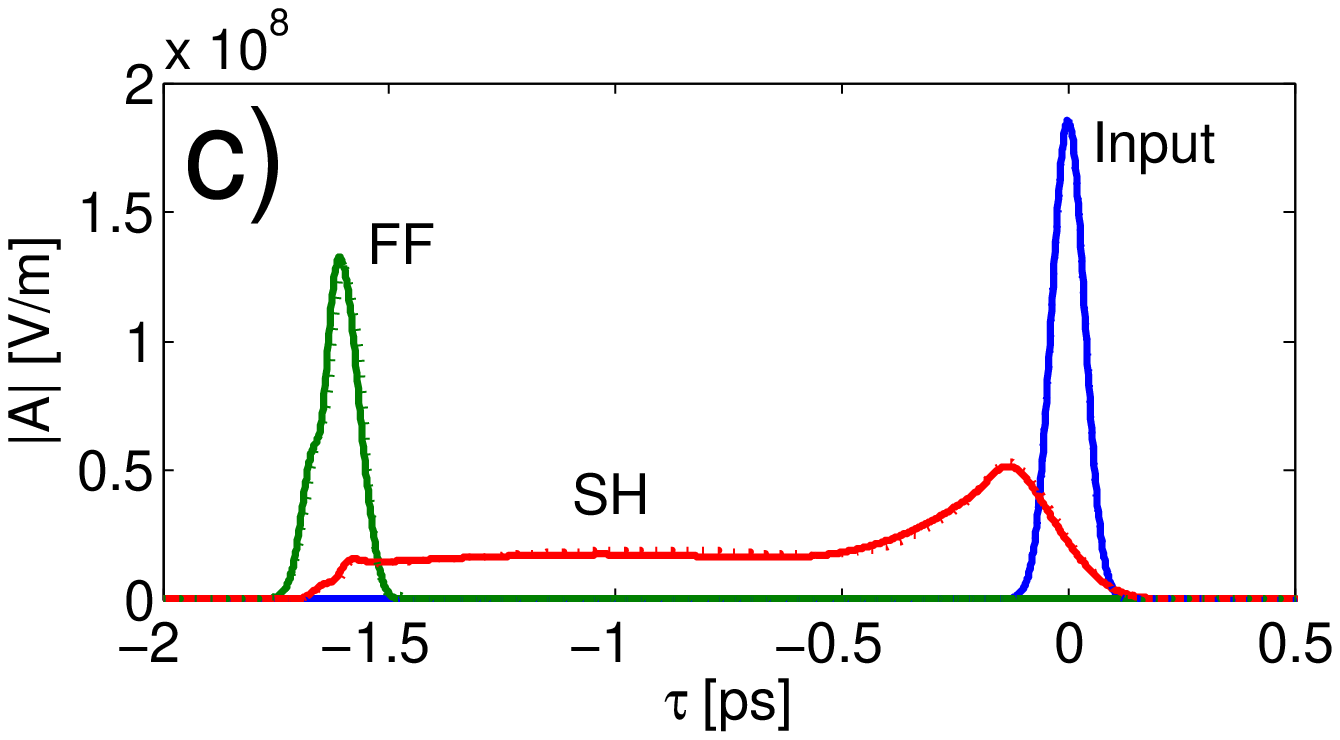}
  \caption{Propagation of a femtosecond pulse in a PPLT crystal. a) Evolution
   of the field amplitude $|A|$ from numerical solution of Eq. (\ref{NEE2}).
   b) Electric field amplitude at the crystal output.
   c) Comparison between coupled wave solution (dotted curves) and
   $|A|$    filtered around fundamental and second harmonic (solid curves).
   The initial pulse has gaussian shape and the parameters are $T=60 fs$,
    $I=10GW/cm^2$, $\lambda_{in}=1400nm$, $\lambda_0=2\pi c/\omega_0=700nm$, $d_{33}=\chi_{LT}^{(2)}/2=10.6 pm/V$ }\label{cfr2onde}
\end{figure}

In order to show the validity of our equation, we simulated the
propagation of a femtosecond pulse in a $L=5mm$ long periodically
poled lithium tantalate sample (PPLT). To model the refractive
index dispersion we employed a Sellmeier model fitted from
experimental data \cite{neslt} and nonlinear coefficient is
$d_{33}=\chi_{LT}^{(2)}/2=10.6 pm/V$. In the numerical code we
inserted the exact dispersion relation $k(\omega)$. We assumed a
first order quasi phase matching (QPM) grating, with a period
$\Lambda=17.4 \mu m$. We thus allowed a periodic variation of the
nonlinear coefficent
$\chi^{(2)}$=$\chi^{(2)}(z)=2/\pi\chi^{(2)}_{LT} e^{i2\pi/\Lambda
z }$+c.c..  We injected a $T=60 fs$ FWHM long gaussian pulse,
centered around $1400 nm$, with $I=10 GW/cm^2$ peak intensity. The
corresponding residual phase mismatch is $\Delta k=
2k(\omega_{in})-k(2\omega_{in})=10000m^{-1}$, where $\omega_{in}$
is the carrier frequency of the input pulse. In the simulation we
set the reference frequency $\omega_0$ to be equal to the second
harmonic of the input pulse: in this way the second harmonic is
stationary
in the reference frame $(z',\tau)$.\\
 Figure \ref{cfr2onde}a)
shows the evolution of the electric field envelope amplitude $|A|$
from numerical solution of Eq. (\ref{NEE2}). We can see the
typical scenario of the propagation of femtosecond pulses in
highly group velocity mismatched (GVM) process: the fundamental
frequency (FF) pulse generates its second harmonic (SH) during
propagation, and the generated SH pulse has the typical shape of a
initial peak followed by a long tail, whose duration is fixed by
the product between GVM and crystal length. Figure
\ref{cfr2onde}b) shows the electric field envelope amplitude at
the end of the crystal. It can be seen a peak  that corresponds to
the faster frequency components located around FF, followed by a
long tail that ends with a second lower peak. This long pulse
corresponds to the generated SH components. The SH pulse is
smooth, indicating that no beating with eventual FF components is
present. Whereas in the residual FF pulses centered around
$\tau\approx-1.6ps$, there is a clear fast oscillation, indicating
that FF and SH components are
superimposed.\\
To test the results, we simulated the same set-up with a standard
coupled wave model \cite{Conforti07}, by inserting the values of
first and second order dispersion evaluated at FF and SH. To
compare the results we filtered $A$ around FF and SH. Figure
\ref{cfr2onde}c) shows the electric field amplitudes at the end of
the crystal. The results of the two models are practically
indistinguishable. It is worth noting that this simulation shows
the validity of Eq. (\ref{NEE2}) over a bandwidth of $\omega_0$.

\begin{figure}[h]
  \centering
      \includegraphics[width=0.4\textwidth]{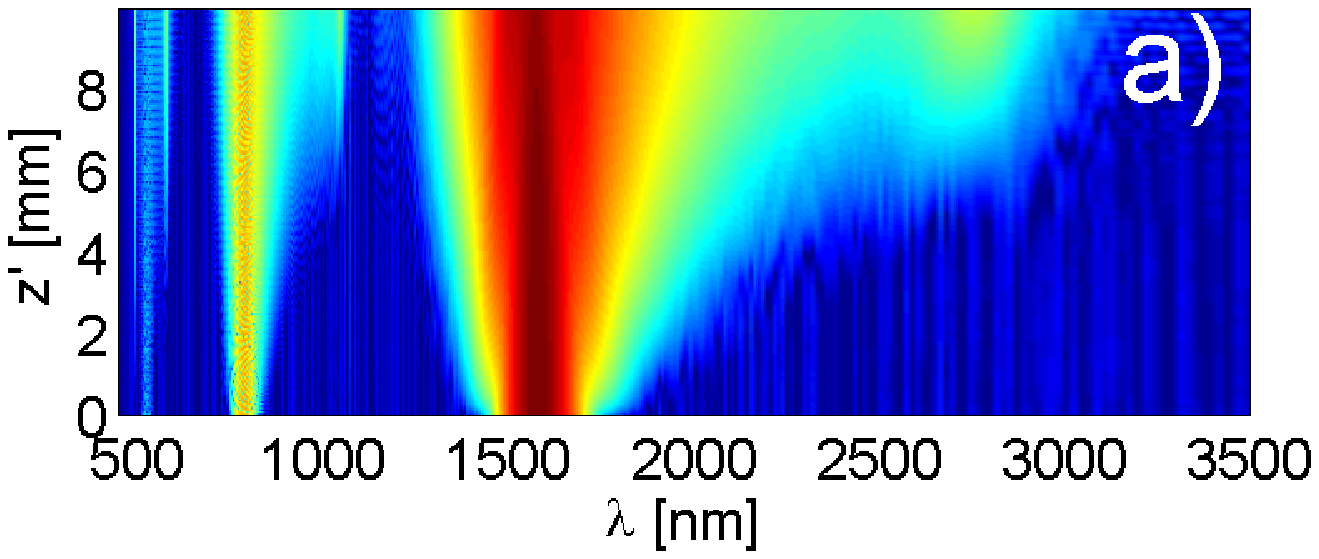}
      \includegraphics[width=0.4\textwidth]{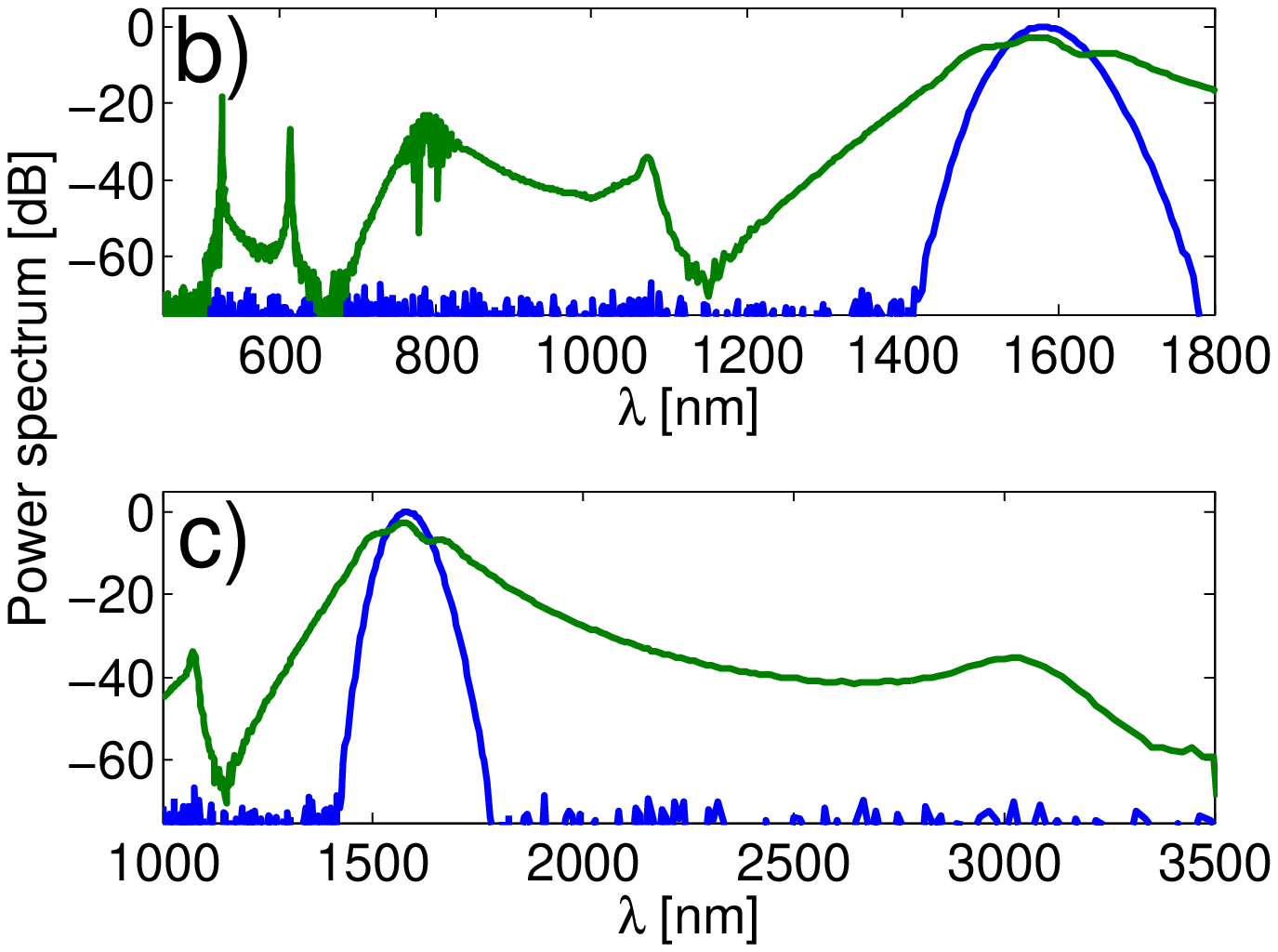}
       \caption{Propagation of a femtosecond pulse into a PPLN sample. a) Evolution
   of the power spectrum (in dB) from numerical solution of Eq. (\ref{NEE2}).
   b) Power specrum at the crystal output in the visible and NIR range.
   c) Power specrum at the crystal output in the infrared range.
   The division into separate spectral region is made to facilitate the comparison with experimental data \cite{Langrock07}.
   The initial pulse has gaussian shape and the parameters are $T=50 fs$,
    $I=15GW/cm^2$, $\lambda_{in}=1580nm$, $\lambda_0=2\pi c/\omega_0=700nm$, $d_{33}=\chi_{LN}^{(2)}/2=27 pm/V$} \label{sc}
\end{figure}

As a second example we consider the propagation of a femtosecond
pulse into a highly mismatched periodically poled lithium niobate
(PPLN) sample, that was demonstrated experimentally to generate an
octave spanning supercontinuum spectral broadening
\cite{Langrock07}.  To model the refractive index dispersion we
employed a Sellmeier model fitted from experimental data
\cite{jundt} and nonlinear coefficient is
$d_{33}=\chi_{LN}^{(2)}/2=27 pm/V$. In the numerical code we
inserted the exact dispersion relation $k(\omega)$. We assumed a
QPM grating with a period $\Lambda=30 \mu m$ (phase matched for
second harmonic generation at around $2 \mu m$ fundamental
wavelength). We included higher order QPM terms, since the huge
bandwidth can phase match different spatial harmonics. We injected
a $T=50 fs$ FWHM long gaussian pulse, centered around $1580 nm$,
with $I=15 GW/cm^2$ peak intensity.  In the simulation we set the
reference wavelength $\lambda_0=700nm$. Figure \ref{sc}a) shows
the evolution of the spectrum during the propagation into a
$L=7mm$ crystal. We can see a consistent broadening  and redshift
of the FF part of the spectrum that, at the end of the crystal,
reaches an octave-spanning bandwidth from $1200nm$ to $3000nm$. We
can also see the generation of spectral components at the second
and third harmonics. At the second harmonic the spectrum initially
broadens and has an evolution ruled by highly mismatched SHG. When
the FF broadening reaches the first order quasi phase matching
wavelength  at around $2\mu m$, the more efficient conversion
process generates a spike at around $1\mu m$. Figure \ref{sc} b)
shows the visible and the near infrared (NIR) part of the spectrum
at the crystal output. We can  see a broadband second and third
harmonic of the broadened laser spectrum, and the presence of some
spikes given by the quasi phase matching of high order spatial
harmonics of the grating. We verified that the two spikes at the
third harmonic correspond to the third and fifth order QPM for the
process $\omega + 2\omega \rightarrow 3\omega$. We can also see a
spectral overlap between the harmonics of the broadened laser
spectrum, that can be exploited to achieve carrier-envelope-offset
phase slip stabilization \cite{Langrock07}, that is of paramount
importance for frequency
metrology applications.\\
Figure \ref{sc}c) shows the infrared spectrum at the output. This
spectrum exhibits more than an octave spanning between $1300 nm$
and $3000 nm$ at the $-40 dB$ spectral power lever with respect to
the peak power level. The spectral components near the zero GVM
wavelength around $3000 nm$ are generated more efficiently.\\
All the features described above compares surprisingly well with
the experimental results of Langrock et al. \cite{Langrock07},
even if we simulate a slightly different environment. In fact we
use a bulk PPLN sample and not a RPE PPLN waveguide. The effect of
waveguide is to slightly modify the power levels and the crystal
dispersion: a detailed simulation of the real set-up is out of the
scope of this Letter. It is worth noting that numerical modelling
of such phenomena without our model is an irksome job since (i)
time domain Maxwell equation solvers require a prohibitive
computational effort and (ii) coupled wave approaches cannot be
used in the presence of overlapping among frequency bands of
different field components.

In conclusion we have derived a robust nonlinear envelope equation
describing the propagation in dispersive quadratic materials.
Thanks to a proper formal definition of the complex envelope, it
is possible to treat pulses of arbitrary frequency content. A
proper definition of envelope is crucial for second order
nonlinearities, due to the generation of frequency components
around zero. Computationally it is possible to accurately evolve
optical pulses of arbitrarily wide band over a meter scale
physical distance, which is a few order of magnitude longer than
those accessible by Maxwell equation solvers.

\


\begin{thebibliography}{}
\bibitem{Boyd} R. W. Boyd, \emph{Nonlinear Optics}, (Academic
Press, 2003), 2nd ed.

\bibitem{Brabec97} T. Brabec and F. Krausz, Phys. Rev. Lett.
{\bf 78}, 3282 (1997).

\bibitem{Brabec00} T. Brabec and F. Krausz, Rev. Mod. Phys.
{\bf 72}, 545 (2000).

\bibitem{Geissler99} M. Geissler et al., Phys. Rev. Lett. {\bf
83}, 2930 (1999).

\bibitem{Housakou01} A. V. Husakou and J. Herrmann, Phys. Rev.
Lett. {\bf 87}, 203901 (2001).

\bibitem{Kolesik02}
M. Kolesik, J. V. Moloney and M. Mlejnek, Phys. Rev. Lett. {\bf
89}, 283902 (2002).

\bibitem{Genty07}
G. Genty, P. Kinsler, B. Kibler and J. M. Dudley, Opt. Express {\bf
15}, 5382 (2007).

\bibitem{Kinsler03}
P. Kinsler and G. H. C. New, Phys. Rev. A {\bf 67}, 023813 (2003).

\bibitem{Moses06}
J. Moses and F. W. Wise, Phys. Rev. Lett {\bf 97}, 073903 (2006).

\bibitem{Langrock07}
C. Langrock, M. M. Fejer, I. Hartl, and M. E. Fermann, Opt. Lett.
{\bf 32}, 2478 (2007).

\bibitem{Haykin}
S. Haykin, \emph{Communication System}, (John Wiley \& Sons,
2001), 4th ed.

\bibitem{neslt}
A. Bruner et al., Opt. Lett. {\bf 28}, 194 (2003).

\bibitem{Conforti07}
M. Conforti, F. Baronio, and C. De Angelis, Opt. Lett. {\bf 32},
1779 (2007).

\bibitem{jundt}
D. H. Jundt, Opt. Lett. {\bf 22}, 1553 (1997).

\end{thebibliography}
\end{document}